# Thermal insulation and heat guiding using nanopatterned MoS$_2$

Peng Xiao[1,2,*,□], Alexandros El Sachat[1,*], Emigdio Chávez Angel[1], Giorgos Nikoulis[3], Joseph Kioseoglou[3], Konstantinos Termentzidis[4], Clivia M. Sotomayor Torres[1,5], and Marianna Sledzinska[1,□]

[1]Catalan Institute of Nanoscience and Nanotechnology (ICN2), CSIC and BIST, Campus UAB, Bellaterra, 08193 Barcelona, Spain
[2]Departamento de Física, Universidad Autónoma de Barcelona, Bellaterra, 08193 Barcelona, Spain
[3]Department of Physics, Aristotle University of Thessaloniki, GR-54124 Thessaloniki, Greece
[4]Univ Lyon, CNRS, INSA Lyon, CETHIL, UMR5008 69621 Villeurbanne, France
[5]ICREA, Passeig Lluis Companys 23, 08010 Barcelona, Spain

[*]These authors contributed equally: Peng Xiao, Alexandros El Sachat.
[□]e-mail: peng.xiao@icn2.cat; marianna.sledzinska@icn2.cat

## Abstract

In the modern electronics overheating is one of the major reasons for device failure. Overheating causes irreversible damage to circuit components and can also lead to fire, explosions, and injuries. Accordingly, in the advent of 2D material-based electronics, an understanding of their thermal properties in addition to their electric ones is crucial to enable efficient transfer of excess heat away from the electronic components.

In this work we propose structures based on free-standing, few-layer, nanopatterned MoS$_2$ that insulate and guide heat in the in-plane direction. We arrive at these designs via a thorough study of the in-plane thermal conductivity as a function of thickness, porosity, and temperature in both pristine and nanopatterned MoS$_2$ membranes. Two-laser Raman thermometry was employed to measure the thermal conductivities of a set of free-standing MoS$_2$ flakes with diameters greater than 20 μm and thicknesses from 5 to 40 nm, resulting in values from 30 to 85 W/mK, respectively. After nanopatterning a square lattice of ~100-nm diameter holes with a focused ion beam we have obtained a greater than 10-fold reduction of the thermal conductivities for the period of 500 nm and values below 1 W/mK for the period of 300 nm. The results were supported by equilibrium molecular dynamic simulations for both pristine and nanopatterned MoS$_2$.

The selective patterning of certain areas results in extremely large difference in thermal conductivities within the same material. Exploitation of this effect enabled for the first time thermal insulation and heat guiding in the few-layer MoS$_2$. The patterned regions act as high thermal resistors: we obtained a thermal resistance of $4·10^{-6}$ m$^2$K/W with only four patterned lattice periods of 300 nm, highlighting the significant potential of MoS$_2$ for thermal management applications.

# Introduction

The modern world is becoming increasingly reliant on wireless and mobile technologies. Maintaining this ever-growing trend requires smaller and more complex electronics that operate at higher frequencies, though this is limited by excessive heat generation during device operation that is exacerbated by the miniaturization of electric components, which often threatens their proper function. Management of this excess heat is a difficult challenge that has resulted in significant recent research efforts to control and manipulate heat transport, which are key issues for thermal engineering and waste heat conversion(*1–6*). A viable solution to these issues could be 2D materials. Materials with very high thermal conductivities (κ) such as graphene (2000 W/mK)(*7, 8*) and hBN (700-450 W/mK)(*9–11*) already find commercial applications in heat spreading. Materials with lower κ, such as $MoS_2$ (25 W/mK -100 W/mK, for single layer and bulk, respectively(*12–16*)) and other semiconducting transition metal dichalcogenides (TMDs) have potential for applications in thermoelectric generation (*17, 18*), thermal rectification(*19*), or thermal insulation and heat guiding. In particular, the last two have significant implications for 2D material based electronics by enabling the precise routing of heat away from hotspots, to avoid damaging subjacent structures. However, such devices have yet to be demonstrated using TMDs.

One reason that TMDs have yet to observe more widespread adoption for these applications is that a *local* control of κ is needed. While the thermal conductivity of $MoS_2$ can be decreased by introducing defects or grain boundaries(*20, 21*), there are no existing systematic studies on engineering the local κ. One possibility to reduce κ is via nanostructuring of the material, such as patterning periodically arranged holes. These structures can be achieved using standard nanofabrication techniques, such as electron beam lithography and dry etching or focused ion beam (FIB). These techniques have been shown to work efficiently for other materials, such as silicon, as a consequence of a reduction of the acoustic phonon mean free paths (MFPs)(*22, 23*). In the case of $MoS_2$, existing theoretical work points to rather short phonon MFPs, on the order of 5–20 nm, which suggests that any nanostructuring should have a comparable periodicity(*12, 15, 24*). However, a recent theoretical study indicated that even larger nanostructures with a periodicity of about 400 nm could already significantly reduce the in-plane κ for monolayer $MoS_2$(*25, 26*).

Basic approaches for the nanostructuring of $MoS_2$ such as atomically-reducing the thickness by plasma or highly anisotropic wet etching of pre-patterned circular holes with atomic precision have been reported recently,(*27–29*) but their effect on κ has yet to be systematically studied. Additionally, there are no studies on the effect of periodicity and disorder, such as defects or amorphisation, on thermal transport in $MoS_2$.

In this work we combined fabrication, measurement, and simulation techniques to understand thermal transport in few-layer pristine and nanopatterned $MoS_2$, and propose thermal management solutions for heat blocking and guiding. We measured the in-plane κ of large-area, free-standing, few-layer $MoS_2$ membranes using two-laser Raman thermometry (2LRT). Subsequently, samples with lattice periods ranging 300 to 500 nm were fabricated in these same membranes using a focused ion beam (FIB), and their thermal conductivity was re-measured. We used the Equilibrium Molecular Dynamics (EMD) approach based on the autocorrelation function of the heat flux to simulate the effect of film thickness and temperature on the thermal conductivity of pristine and patterned MoS₂. Based on the obtained results, we designed and fabricated two thermal management devices: a thermal insulating ring which confines heat in a delimited area, and a heat guiding channel which confines heat and conducts it away from a source. This is the first time that such kind of devices have been successfully realized for 2D materials. These results indicate that nanopatterned MoS₂ is a promising platform for thermal applications.

## Results

**Sample fabrication and characterization**

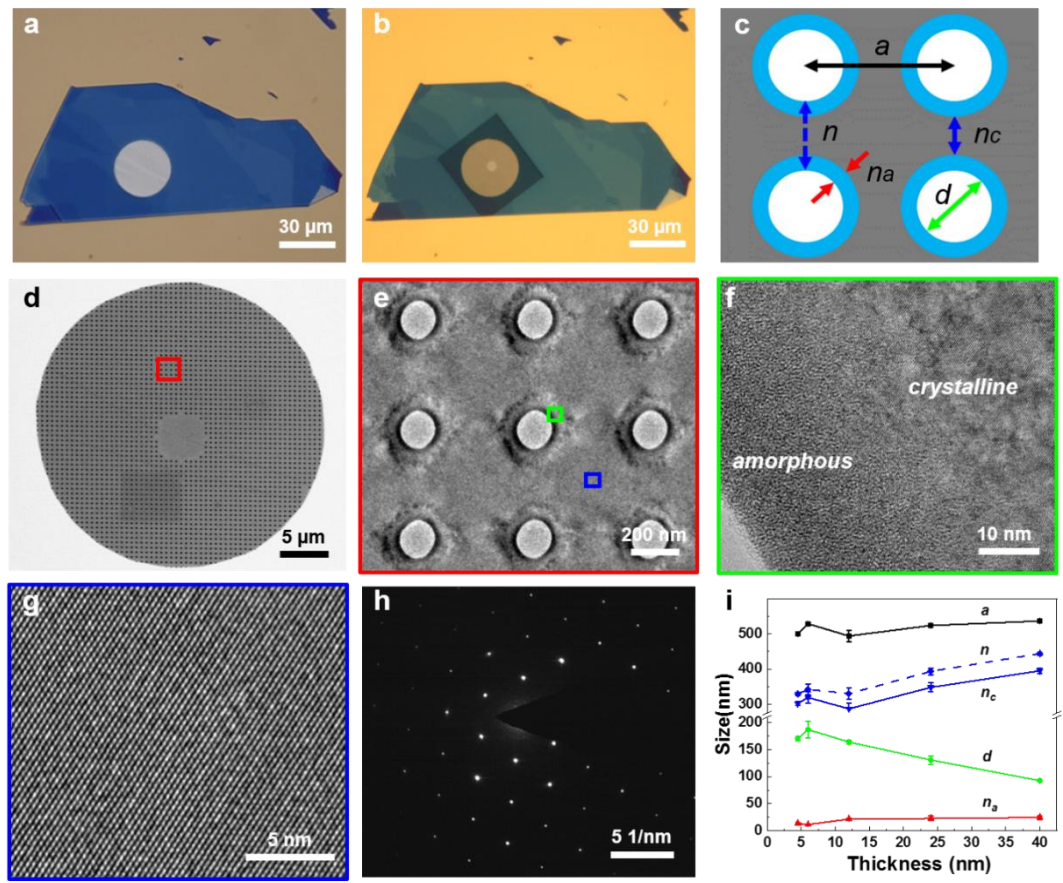

A set of eight free-standing, pristine MoS₂ membranes with thicknesses from 5 to 40 nm and diameters of 20 and 30 μm was fabricated using the exfoliation and dry transfer method

**Fig. 1 Characterization of the 12 nm-thick, free-standing MoS₂.** Optical micrographs of a suspended MoS₂ flake: **a,** pristine and **b,** after FIB nanopatterning. **c,** Schematic of the patterned sample with period $a$, hole diameter $d$, neck $n$, crystalline neck $n_c$, amorphous neck $n_a$, where $n = a-d = n_c + n_a$. **d,** SEM image of the patterned sample; **e,** TEM image, at the region that corresponds to the red square in panel (**d**). **f,** High resolution TEM image of the hole edge at the region that corresponds to the green square in panel (**e**). **g,** High resolution TEM image of the crystalline neck at the region that corresponds to the blue square in panel (**e**). **h,** Corresponding electron diffraction pattern **i,** geometrical parameters as a function of the MoS₂ thickness.

(*Materials and Methods* and Fig. S1). An example of a free-standing, 12 nm-thick $MoS_2$ membrane with diameter of 30 µm is shown in Fig. 1a. FIB was used in order to create periodic holes through the membranewith a hole lattice period period *a* of 500 nm (Fig. 1c). The central area was intentionally left unpatterned to be used to focus the heating and probing lasers. Figure 1b-e shows optical and scanning electron microscopy (SEM) images of the same 12 nm-thick $MoS_2$ membrane after nanopatterning.

FIB patterning is known to damage crystalline structure and introduce amorphisation in processed samples, which has strong implications on thermal transport(*30*). To minimize the damage, a low gallium ion current (2 pA) was used in this work. In the TEM, an amorphous region with width $n_a$ of approx. 10 – 30 nm can be seen around each hole. The interface between the amorphous and crystalline phase is not sharp, the two phases are mixed, as can be seen in Fig. 1f. The size of the amorphous region around the holes patterned using the FIB is comparable to other fabrication techniques, such as laser ablation (Fig. S4). The remaining area between the holes $n_c$ maintained its crystalline structure, as confirmed by TEM and electron diffraction (Fig.1g and h). A summary of the dimensional parameters of all the samples and corresponding TEM images are shown in Fig. 1i and Fig. S2. A very small presence of Ga ions resulting from the FIB patterning was detected only in the 40 nm-thick sample (Fig. S3) and thus is not expected to significantly affect the thermal transport.

**Thermal conductivity measurements**

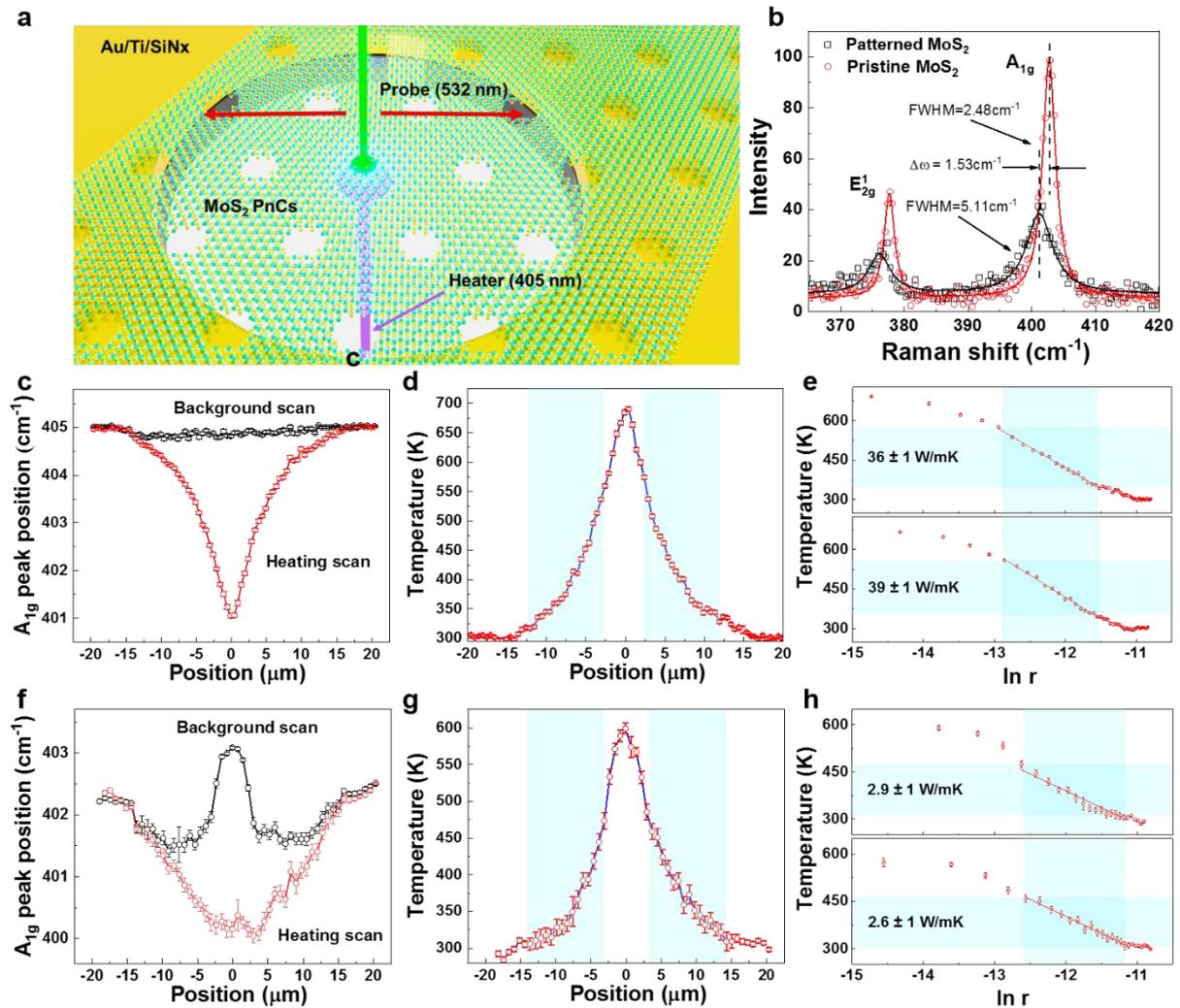

**Fig. 2 Thermal conductivity of the 12nm-thick, free-standing MoS$_2$. a,** Schematic of the 2LRT set-up. **b,** Raman spectra of the pristine and patterned MoS$_2$ membrane. **c-e** The thermal conductivity results of the MoS$_2$ pristine membrane. **c,** *Background and Heating scans* were acquired with the probe laser with the incident power with $P_{probe}$ ~10 µW and the heating laser power absorbed by the sample $P = 0$ µW and $P = 157$ µW, respectively. and the probe laser. *scan* corresponds to the heating laser with and $P_{probe}$ ~10 µW. The heating laser is focused onto the center of the sample ($r = 0$ µm). **d,** Temperature profile on the sample, extracted from **(c)**. **e,** Temperature profile as a function of ln($r$). The solid line represents a linear fit of the experimental points. **f-h** The results of the MoS$_2$ patterned membrane. **f,** *Background scan* corresponds to the heating laser with $P = 0$ µW and the probe laser with $P_{probe}$ ~6.8 µW. *Heating scan* corresponds to the heating laser power absorbed by the sample with $P = 17.5$ µW and $P_{probe}$ ~6.8 µW. **g,** Temperature profile on the sample, extracted from **(f)**. **h,** Temperature profile as a function of ln($r$).

To measure the thermal conductivity of the pristine and patterned MoS$_2$ membranes, 2LRT was used in the configuration shown schematically in Fig. 2a and described in detail in the *Materials and Methods* section. The heating laser (405 nm) coupled to the sample from the bottom creates a temperature gradient which translates to a shift in the frequency of the Raman active modes of MoS$_2$, which are measured using the low-power probe laser (532 nm) coupled to the Raman spectrometer.

Thermal transport in 2D free-standing membranes is dominated by the in-plane thermal conductivity, which can be calculated using Fourier's law: $P/(2\pi rt) = -\kappa\, dT/dr$ where $P$ is the power absorbed by the membrane, $r$ is the distance from center, and $t$ is the membrane thickness. By taking $rdT/dr = dT/d(lnr)$ the following expression for κ was obtained: $\kappa = -P/(2\pi t \frac{dT}{d(lnr)})$. Due to the system symmetry, in 2LRT the line scan Raman spectra can be collected when the probe laser linearly scans the free-standing membrane through the center (Fig. 2c-h) (*20, 23, 31*).

An example of the experimental procedure is shown in Fig. 2 for the 12-nm thick pristine MoS$_2$ membrane from Fig. 1. First, the background scan was performed with the heating laser $P = 0$ μW and the probe laser set to a low power ($P_{probe}$ ~10 μW). Next, the heating scan was performed with the heating laser power absorbed by the sample $P = 157$ μW. The A$_{1g}$ peak position for both of the scans is shown in Fig. 2a. The background scan curve shows the constant peak position at 405cm$^{-1}$, indicating no significant strain is present in the sample. The heating scan curve shows symmetric temperature decay from the center of the samples to the heatsink, where it reaches room temperature. The A$_{1g}$ peak position difference between the background scan and heating scan was divided by the A$_{1g}$ peak's temperature coefficient of each sample, and converted to the temperature (Fig. 2d, S5 and Table T1). Near the heat-sink, the profiles could be well fitted by the constant κ, as shown in Fig. 2e, where the slope of the fitting lines in corresponds to the $dT/d(lnr)$.

After nanopatterning, three interesting features were revealed in the Raman spectra of the MoS$_2$ (Fig. 2b, f): (i) a red-shift of the Raman modes, (ii) broadening of their linewidths, and (iii) a remarkable decrease in their intensities. The modification of the Raman spectra was a clear indication of the presence of the amorphous phase in the patterned membranes. The absence of spatial order and long-range translational symmetry led to red-shifting of the Raman modes and a broadening of their linewidths(*32*).

The patterned membranes were measured in the same way as their pristine counterparts. For the patterned 12 nm-thick MoS$_2$ membrane the intensity of the probe laser was reduced to 6.8 μW, while the heating laser power absorbed by the sample $P$ was 17.5 μW (Fig. 2f-h). To extract the intrinsic thermal conductivity of the patterned MoS$_2$, the experimental value was corrected using a volume correction factor ε, which takes into account the volume reduction of the patterned samples, where $\varepsilon = (1-\sigma)/(1+\sigma)$ and the porosity σ = $\pi d^2/4a^2$ (*33*).

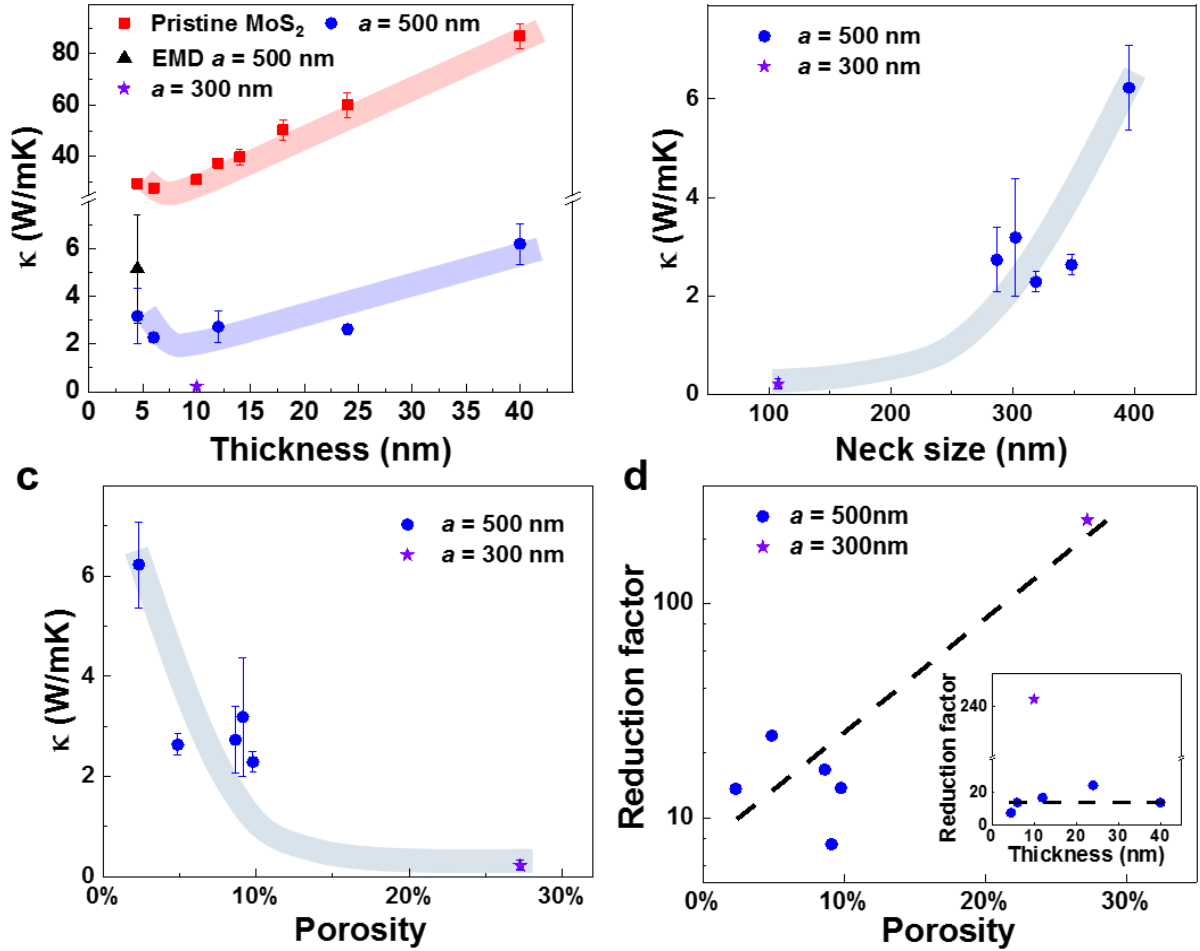

**Fig. 3 Thermal conductivity of the pristine and patterned MoS$_2$. a,** Comparison between the thermal conductivities of the pristine and nanopatterned MoS$_2$ measured using 2LRT and simulated using EMD. **b-c,** Thermal conductivity of the nanopatterned MoS$_2$ as a function of the neck size and porosity, respectively. **d,** Thermal conductivity reduction factor as a function of porosity and thickness (inset). Lines serve as guides to the eye.

Thermal conductivity as a function of the thickness of MoS$_2$ is shown in Fig. 3a. The κ of pristine MoS$_2$ increased from 34 ± 3 W/mK to 88 ± 5 W/mK as the thickness increased from 6 nm to 40 nm. We observed almost linear increase of the κ as a function of thickness. A small minimum observed for the 6 nm thick sample can be explained as competition between the phonon-phonon and surface-phonon scattering (*34*). This result, previously observed also for supported MoS$_2$, was reported at around 7 layers, in agreement with the results in this work.

In comparison, the patterned membranes with *a* = 500 nm showed a significant, 10-fold reduction of thermal conductivity. The patterned membranes followed the same trend as the pristine ones, with a small dip in the thermal conductivity for the 6 nm-thick sample. An additional sample with *a* = 300 nm showed an even lower thermal conductivity, below 1 W/mK which approaches the amorphous value(*14*). In Fig. 3 b and c we compared the κ as functions of the neck size and porosity. For *a* = 500 nm κ decreased almost linearly with neck size and porosity. However, for $a$ = 300 nm we observed a sharp drop in κ, which might be similar to the effect observed previously for patterned silicon membranes(*35*).

Finally, to evaluate the impact of nanopatterning on κ, a reduction factor $R$ was defined as $R = (\kappa_p - \kappa)/\kappa$, where $\kappa_p$ and $\kappa$ are thermal conductivities of the pristine and patterned MoS$_2$, respectively. As shown in Fig. 3d, $R$ increased from ~8 to ~244 when the porosity increased from 0.09 to 0.274. At the same time, for period of 500 nm and varying thickness R varied between 10 and 20. This further confirms the predominant role of the hole boundary scattering. Such large $R$ indicates that the nanopatterned MoS$_2$ is a promising candidate for heat blocking.

## Discussion

**NEMD simulations of pristine and patterned MoS$_2$**

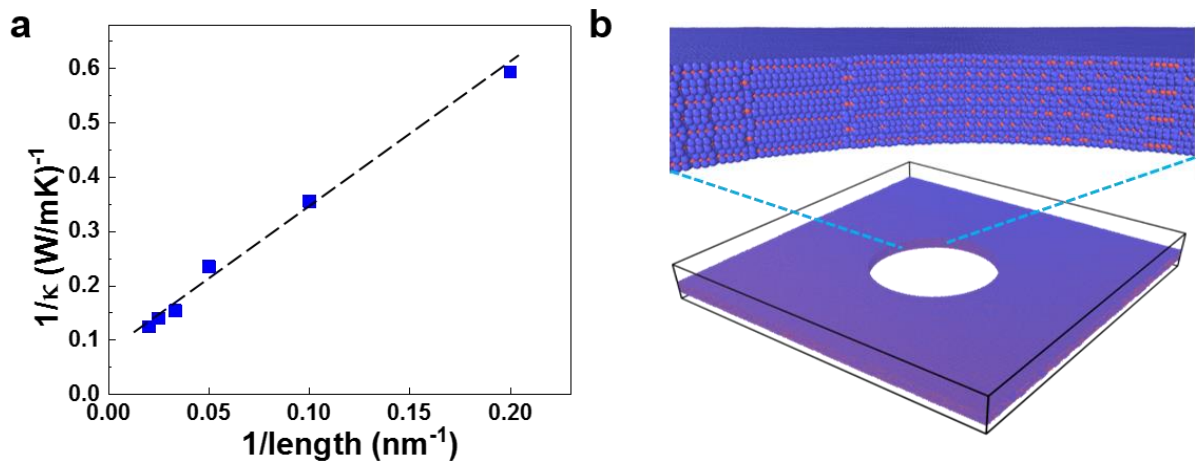

**Fig. 4 a,** System size dependence of 1/κ leading to the calculation of the phonon mean free path at 300 K for a 4 nm-thick MoS$_2$ membrane. **b,** Atomistic configuration of the 4 nm-thick MoS$_2$ nanopatterned membrane containing 2,037,703 atoms; 97.32 nm along x direction and 97.14 nm along y direction. Inset: surface atoms of the hole.

We used MD simulations to better understand the thickness and patterning effects on the thermal conductivity of MoS$_2$ (*Materials and Methods*). The thermal conductivity was studied with the EMD method(*36*) using the LAMMPS package and the REBO-LJ potential(*37*). In previous studies using the REBO-LJ potential and the homogeneous non-equilibrium MD the highest κ value was found for the single-layer MoS$_2$ (100 W/mK), which further decreased with increasing thickness and reached the bulk value for three layers(*38*). In the current study, using the EMD method, we found that the in-plane thermal conductivity increased from ~9 W/mK to 24 W/mK, for the 4- and 10-nm thick MoS$_2$, respectively. For the 10 nm thick MoS$_2$, its thermal conductivity already reached 62% of the bulk value. The increase in κ as a function of the MoS$_2$ thickness obtained with the EMD simulations was in good agreement with the experimental results (Fig. S7).

To calculate the phonon MFPs in the in-plane direction of the 4 nm thick $MoS_2$, non-equilibrium MD (NEMD) simulations were performed following the methodology described in Ref[9]. The inverse thermal conductivity versus the inverse length of the atomistic model showed a linear dependence (Fig. 4a), from which the MFP was calculated to be about 41 nm.

A similar value of 30 nm for bulk $MoS_2$ was obtained using the MFP reconstruction method. The reconstruction method was applied to the experimental results assuming diffusive thermal transport governed by the Fuchs-Sondheimer approach (Fig. S6) (*39, 40*). An extended description of this method can be found elsewhere (*41–44*).

The results of the phonon MFP calculations were used to scale down the atomistic model of the nanopatterned samples that have been investigated experimentally. The thickness of the atomistic model of the $MoS_2$ was kept at 4 nm, while the dimensions along the x and y directions were scaled down five times. Regardless, the distance between two consecutive holes was larger than the average phonon MFP (Table T3). Thus, we could study the effect of nanostructuring of the membranes without large error by scaling down the model structure. It should be noted that the full-sized atomistic model would contain about 60,000,000 atoms and our initial calculations concluded that a study on such a model would have been inaccurate and computationally inefficient.

In Fig. 4b the atomistic configuration of the unit cell of the nanopatterned $MoS_2$ is presented(*45*). The inset shows a zoomed-in region on the surface atoms of the hole. The sulfur atoms are irregularly displayed on the hole surface, due to their higher mobility with respect to the molybdenum atoms and the hole's surface is partially covered with sulfur atoms.

The 4 nm-thick $MoS_2$ membranes with $a = 500$ nm showed an experimental thermal conductivity of about $3.5 \pm 1$ W/mK. For the scaled down atomistic model from Fig. 4b, the in-plane thermal conductivity calculated by EMD is $5.2 \pm 2.3$ W/mK, a value that is in agreement with the experimental values, and hence validates the aforementioned modeling procedure. The difference between the experimental and theoretical value can be attributed to the presence of the amorphous ring around the perforated holes. In previous studies, the presence of a native oxide or amorphous phase above the nano-membranes or on the walls of the holes in the patterned nanomembranes were shown to have significant impact on the effective thermal conductivity of such nanostructures(*46–48*). For the case of patterned membranes, the reduction due to the presence of an amorphous shell around the holes has been estimated to be on the order of 50-75%(*47*).

**Temperature-dependent thermal conductivity**

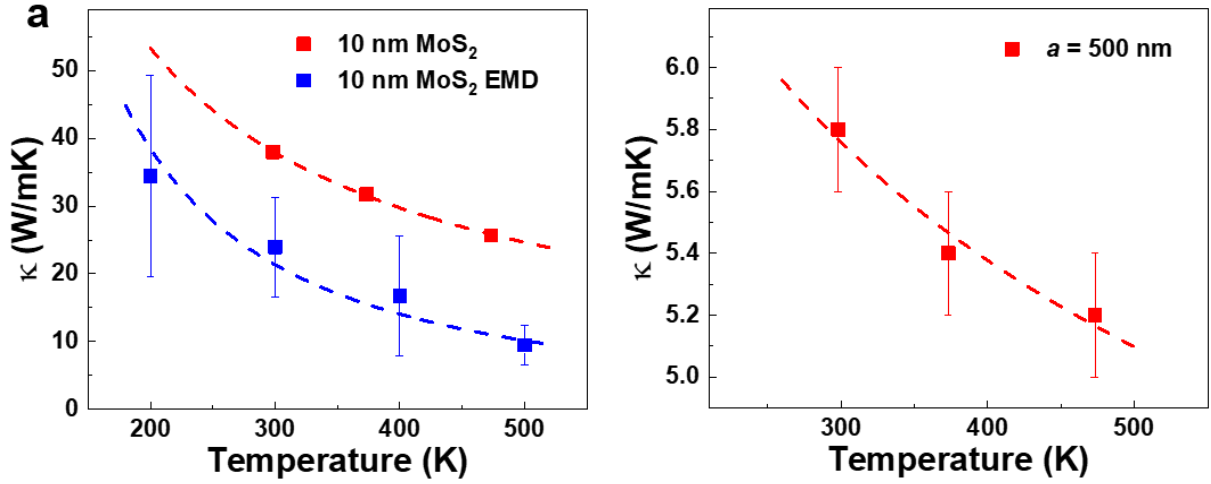

**Fig. 5 Temperature-dependent thermal conductivity of the 12 nm-thick, free-standing MoS$_2$. a,** Comparison between experiment and EMD simulations of temperature-dependent thermal conductivity of a free-standing 12-nm thick MoS$_2$. **b,** Experimental temperature-dependent thermal conductivity of a patterned MoS$_2$. Dashed lines are guides to the eye.

For the high-temperature applications it is crucial to know the temperature dependence of the thermal conductivity κ(T) for both the pristine and patterned MoS$_2$ membranes. An example of the results for the 12 nm-thick MoS$_2$ are shown in Fig. 5 and S8. The calculations were only performed for the pristine membranes due to the aforementioned computational restrictions. For the pristine membrane a 30% reduction in κ was measured as temperature was increased from 300 to 450 K, similar to the decrease predicted by the EMD. For the patterned MoS$_2$, a reduction of about 10% was measured for the same temperature increase. These results confirmed that the temperature more strongly affects the phonon MFPs of the pristine membrane than that of the patterned MoS$_2$. For the pristine MoS$_2$, Umklapp phonon-phonon and higher order scattering are dominant at higher temperatures. On the contrary, for the patterned MoS$_2$, scattering from the holes constitutes the majority of scattering events and thus phonon-phonon interactions are less significant. Even so, at 500 K the five-times difference in κ between the patterned and pristine MoS$_2$ was observed, validating its potential for applications at high temperatures.

**MoS$_2$ thermal insulator and heat conduction channel**

The large difference between the pristine and patterned MoS$_2$ thermal conductivities over a wide temperature range can be employed for various thermal management applications. Here, we explored two of them, namely: thermal insulation and heat guiding. As shown in Figs. 6a and S9, a heat insulating ring ($r$ =7.5 μm, $a$ = 300 nm) with four lattice periods was fabricated on a 10 nm thick free-standing MoS$_2$ membrane. As all the measurements were performed in vacuum (no convection or heat transfer to air), no other dissipation channels were available and the only path for the heat to evacuate was through the patterned area.

The heating laser was focused onto the center of the membrane and the temperature profile for the absorbed heating power $P = 28.4$ µW was recorded using the 2LRT set-up, as indicated by the dashed line in Fig. 6a. The center of the membrane, i.e. inside the ring, maintained a temperature between 535 and 480 K. A sharp drop of the temperature from 480 to 330 K was recorded between the inner and outer parts of the ring (highlighted in Fig. 6b), meaning that it blocked most of the heat flow from the center to the edges of the membrane. The temperature of the outer part of the $MoS_2$ was almost unaffected by the heating and maintained a temperature between 330 and 300 K.

The corresponding thermal resistance $R$ of the patterned region can be calculated from $Q = \Delta T/R$, where $Q$ is the heat flux and $\Delta T$ is the temperature drop. The calculated $R$ is $4 \cdot 10^{-6}$ $m^2K/W$, which corresponds to the thermal boundary conductance $G = 1/R = 0.25$ $MW/m^2K$ and thermal conductivity below 1 W/mK.

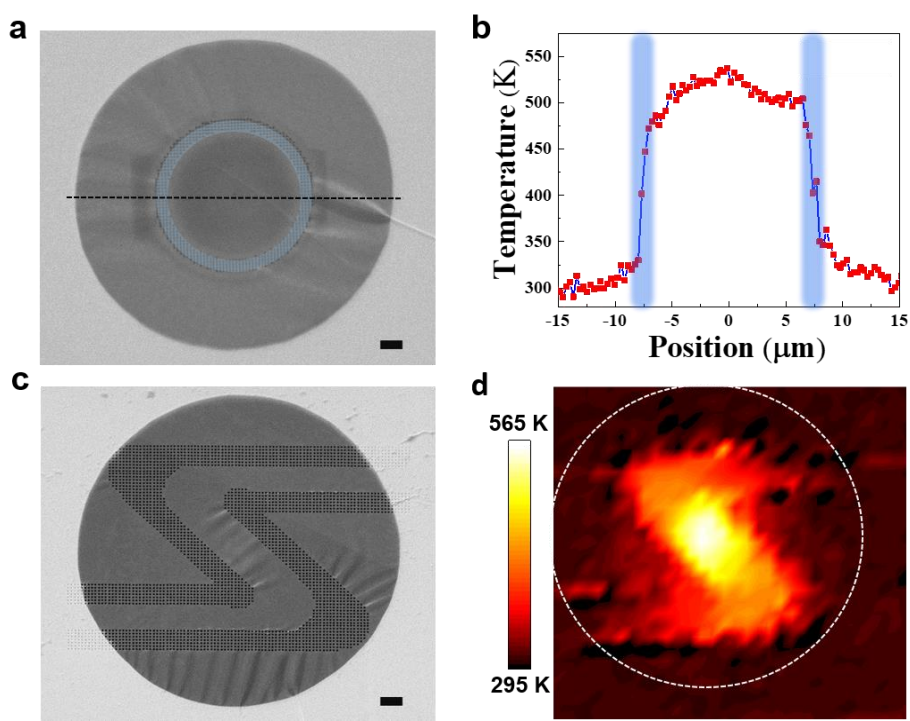

**Fig. 6 $MoS_2$ thermal insulator and heat conduction channel. a,** SEM image of a free-standing 10nm-thick $MoS_2$ with a thermal insulator ring. Dashed line indicates the scan direction. **b,** Corresponding temperature profile for the absorbed power $P = 28.4$ µW focused on the center of the sample. The highlighted areas correspond to the thermal insulating ring. **c,** SEM image of a Z-shape heat conduction channel which was fabricated on a free-standing 10nm thick $MoS_2$ membrane. **d,** Corresponding temperature map for the absorbed power $P = 8.5$ µW focused in the center of the sample. Dashed line corresponds to the membrane edge.

These results were employed to fabricate a Z-shaped heat guiding channel ($a = 300$ nm) on 10-nm thick MoS$_2$ membrane with 30 μm diameter (Figs. 6c and S10). The membrane was heated up to 560 K in the center and a temperature map was recorded using the 2LRT set-up. Figure 6d shows the temperature distribution inside the channel, decreasing from the hotspot towards the heatsink at the membrane edges. Clearly, the lateral heat flow was delimited by the patterned area. These results demonstrate the efficiency of the nanopatterning of the TMDs for heat management applications.

## Conclusions

In this work we have successfully demonstrated local modification of MoS$_2$ thermal conductivity by means of FIB nanopatterning. We have obtained a more than 10-fold reduction of κ for holes with a period of 500 nm, easily accessible with standard nanofabrication tools. A thermal conductivity below 1W/mK was reached with a period of 300 nm. The results were confirmed by the EMD simulations for the pristine membranes and on an adequately reduced unit cell for the nanopatterned membranes. We have proven that even though the phonon MFPs are on the order of tens of nms, the κ is strongly suppressed by structures with periodicities of a few hundreds of nms. The study of the effect of the temperature on κ showed that the main scattering process for pristine membranes is the Umklapp phonon-phonon scattering, while for the perforated membranes scattering along the phonon-hole-walls becomes more important.

The large difference in thermal conductivity between the nanopatterned and pristine membranes, which is maintained even at high temperatures, motivated the design of thermal insulators or heat conduction channels in the MoS$_2$ membranes. The proposed designs pave the way for efficient heat manipulation using 2D materials, with possible applications in 2D electronics for IoT or nanoscale opto- or electro-mechanical free-standing systems.

## Materials and Methods

**Sample preparation and characterization**

A Au/Ti (95 nm/5 nm) layer was deposited on a holey SiNx substrate (Norcada, Canada) by e-beam metal deposition. A 2-mm thick Polydimethylsiloxane (PDMS) film was created using a 10 to 1 ratio of silicon base to curing agent (Sylgard ®184, Dow Corning, USA) and cured at room temperature for 24 hours. The PDMS film was used to mechanically exfoliate $MoS_2$ membranes from bulk $MoS_2$ crystals (Graphene Supermarket, USA), and the $MoS_2$ membranes was dry-transferred onto the Au/Ti coated holey substrate as the free-standing $MoS_2$ membrane.

A Focused Ion Beam (FIB) (Zeiss 1560XB Cross Beam, Gremany) was used to etch periodic holes on the $MoS_2$ membrane with a beam current of 2 pA, a voltage of 30 kV, and an etch time of 10 ms. The holes were fabricated on the free-standing membranes using FIB leaving an unetched center area with a diameter of ~5 μm to be used as a heating island.

**Thermal conductivity measurements**

*Two-laser Raman Thermometry*

As the peak position of a Raman peak depends on the material temperature, it can be used to probe the material temperature. The $A_{1g}$ peak position of $MoS_2$ Raman spectrum was used to probe the temperature of $MoS_2$ membrane in this work, because its peak intensity was much stronger than the $E_{2g}^1$ peak.

A probe laser (532 nm, Cobolt) and a heating laser (405 nm, Cobolt) were focused on the center of a free-standing membrane from the top-side and bottom-side, respectively. The heating laser and the free-standing membrane were fixed on a motorized stage (Marzhauser), then their relative position and the focus of the violet laser remained unchanged during the measurement to ensure a stable and uncontacted heating source. The probe laser was coupled to the Raman spectrometer (T64000) and scanned the sample collecting a Raman spectrum at various points. The probe laser spot size was 1.2 μm. Laser power absorbed by the membranes was measured *in-situ* using the configuration described elsewhere(*20*, *23*).

All the samples were measured in a temperature-controlled, vacuum chamber at a pressure of ~ $3 \times 10^{-3}$ mTorr (Linkam). The gold layer also acted as a heat sink to ensure that the $MoS_2$ temperature in the supported area is the same as that of the environment. For calculating the thermal conductivity, a temperature distribution around the heating source on the free-standing membrane was required.

The 2LRT experiment consists of two consecutive scans:

1) No heating applied ('baseline'). This measurement also helps to assess the sample quality (strain, contamination, etc.).
2) Heating applied using 405 nm wavelength laser coupled from below the sample.

The $A_{1g}$ peak position difference between the background scan and heating scan was divided by the $A_{1g}$ peak's temperature coefficient of each sample, and converted to the temperature (Fig. S5 and Table T1). The profiles of regions near the heat-sink could be well fitted by the constant κ (Fig. 2e) where the slope of the fitting lines corresponds to the $dT/d\,(lnr)$.

The spectra are collected every 0.5 µm using a Märzhäuser stage with Tango controller, which provides repeatability < 1 µm (bidirectional) and resolution of 0.01 µm (smallest step size).

*One-laser Raman Thermometry*

A heating-probe laser (532nm, Cobolt) was used to heat the samples at the center as well as collect the corresponding Raman spectra. All the measurements were performed in a temperature-controlled vacuum chamber (Linkam) and the laser power absorbed by the samples was measured *in-situ* using the configuration described elsewhere(*21*).

*Temperature-dependent thermal conductivity*

The measurements were performed in a temperature-controlled vacuum chamber (Linkam) where the heatsink temperature was varied from 123 K to 473 K. The samples were characterized using the 1LRT.

**Molecular dynamics simulations**

*EMD*

The thermal conductivity was investigated using the equilibrium molecular dynamics (EMD) method(*36*) and the Non-equilibrium Molecular Dynamics (NEMD). All simulations were conducted using LAMMPS (*37*) and the REBO-LJ potential. The $MoS_2$ parameterization of the REBO-LJ potential was considered from ref (*49*), while the Lennard-Jones parameters for the S-S pair was adapted from ref (*50*) as it was motivated by room-temperature (300 K) calculations whereas ref (*49*) was motivated by zero-temperature calculations. The REBO-LJ potential for $MoS_2$ has been extensively tested on thermal property calculations in comparison with all the available interatomic potentials and has been shown to be the most reliable one (*38*). Several systems are modeled with various thicknesses and temperatures. For each specific atomic model and temperature, the thermal conductivity was calculated by averaging 10 different EMD runs with 10 different initialization seeds for random velocity distributions. The first step was to equilibrate the system with an NVT run for 400 ps at the temperature under investigation. This step is to heat up the system prior to the EMD run to reach the linear (or equilibrated) region faster during the EMD. For the simulations under REBO-LJ potential, a time step of 0.5 fs was used and the EMD method for 1200 ps (2400000 steps) was performed. Finally, the average of 10 different cases, each of them having different initial distribution of atomic velocities, was considered to calculate the mean curve of the thermal conductivity vs time plot.

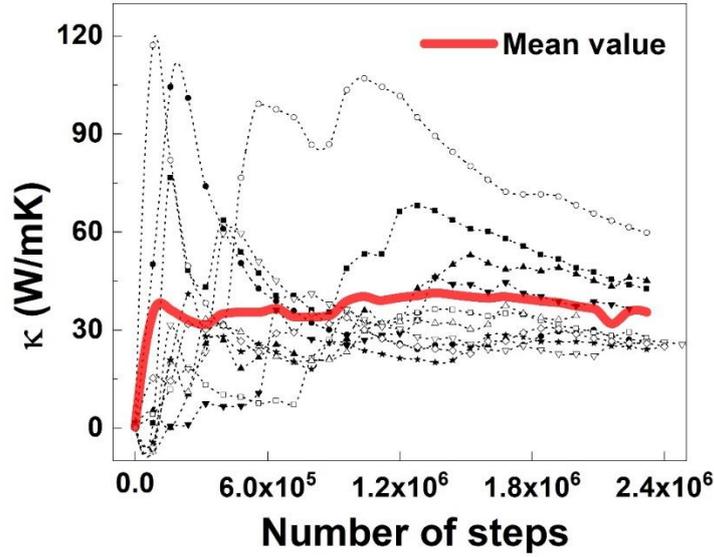

**Fig. 7** EMD simulation for 10nm pristine film at 200K temperature. The plot shows the in-plane results for y-direction (010) and the mean value was calculated using 10 random seeds.

*NEMD*

In order to calculate the phonon MFP of the 4-nm thick 2D $MoS_2$, non-equilibrium Molecular Dynamics (NEMD) simulations at 300 K were performed. 2D atomistic configurations having length in a range from 5 to 50 nm were considered and their thermal conductivity was calculated. The two fixed regions and two thermostat regions have a length of 0.5 nm each and the two thermostat regions are considered in the aforementioned lengths. Following NEMD, a phonon is travelling in the crystal between the two cold and hot reservoirs at temperatures of 270K and 330K, respectively. It is known that phonons can exhibit at least two kinds of scattering procedures: they can be scattered by other phonons travelling in the material, or they can be scattered by the reservoirs, which are considered by the phonon as a different material with an almost infinite thermal conductivity.

# Acknowledgements

This work has been supported by the Severo Ochoa program, the Spanish Research Agency (AEI, grant no. SEV-2017-0706) and the CERCA Programme/Generalitat de Catalunya. MS, ECA and CMST acknowledge support from Spanish MICINN project SIP (PGC2018-101743-B-I00), H2020-FET project NANOPOLY (GA No. 289061), ERC-ADG project LEIT (GA No. 885689). P.X. acknowledges support by a Ph.D. fellowship from the EU Marie Skłodowska-Curie COFUND PREBIST project (GA No. 754558). A.E.S acknowledges support by the H2020-MSCA-IF project THERMIC (GA No. 101029727). The work was supported by computational time granted from the National Infrastructures for Research and Technology S.A. (GRNET) in the National HPC facility - ARIS - under the project NOUS (pr010034).

# Author contributions

P.X. M.S. A.E.S. C.M.S.T conceived the project. P.X. performed sample fabrication. P.X. A.E.S E.C.A. conducted the thermal conductivity measurements under the supervision of M.S. and C.M.S.T. K.T. J.K. G.N. performed the theoretical simulation. P.X. A.E.S. E.C.A. compiled and analyzed the data, with input from K.T. J.K. G.N. developed the model that simulated the experiments.  P.X. and M.S. wrote the paper, with input from all authors.

# Competing interests

The authors declare no competing interests.